# Impacts of photocatalytic hydrogen production on the European energy system

Wolfram Tuschewitzki[1*], Jelto Lange[1], Jacob Schneidewind[2,3,4], Martin Kaltschmitt[1]

[1]Institute of Environmental Engineering and Energy Economics (IUE),
Hamburg University of Technology (TUHH), Eissendorfer Str. 40, 21073 Hamburg, Germany

[2]Institute of Organic and Macromolecular Chemistry (IOMC),
Friedrich Schiller University Jena, Humboldtstr. 10, 07743 Jena, Germany

[3]Center for Energy and Environmental Chemistry Jena (CEEC Jena),
Friedrich Schiller University Jena, Philosophenweg 7a, 07743 Jena, Germany

[4]Helmholtz Institute for Polymers in Energy Application Jena (HIPOLE Jena),
Lessingstraße 12-14, 07743 Jena, Germany

*Corresponding author(s). E-mail(s): wolfram.tuschewitzki@tuhh.de

**Abstract**

Especially in regions with high solar irradiation, photocatalysis presents a promising low-cost "green" hydrogen production option. Thus, this paper analyzes impacts of increasing photocatalysis shares on the European energy system using an open-source energy system optimization model covering the electricity, industry, and heating sectors with high spatial and temporal resolution. Photocatalysis deployment is investigated at various market shares by exogenously altering photocatalysis costs. The results show that integrating photocatalysis necessitates systematic adjustments since it lacks the flexible load attributes of water electrolysis. Therefore, a significant geographic shift in hydrogen production and demand from the Northwest to South Europe is expected in the case of large-scale photocatalysis adoption. Despite these challenges, installed photocatalysis shows costs within the photocatalysis cost projections. Thus, photocatalysis could contribute to a critical diversification of hydrogen production, easing material demands for other renewable technologies. Nevertheless, it requires strategic planning to avoid lock-ins and to maximize its potential.

# Main

To mitigate anthropogenic climate change and associated global warming, resulting in increasingly frequent and severe weather extremes (e.g., heatwaves, droughts, wildfires, extreme precipitation), human-generated greenhouse gas emissions must be curbed [1]. Achieving this requires a fundamental transition of the global energy system from its current reliance on fossil resources to renewable sources of energy. In "renewable" energy systems, hydrogen plays a pivotal role in sector coupling [2], applied in the transport sector, as an industrial feedstock, as an energy carrier for the heat market [3] and enabling seasonal energy storage, thus enhancing system security.

There are two primary methods for producing renewable energy-based "green" hydrogen: water splitting and biomass utilization. The latter is limited by land availability and competition with the production of high-value products (e.g., food) [4]. Additionally, biomass is a valuable carrier for "green" carbon. These arguments make water splitting with electricity (electrolysis) the most widely and often only discussed scalable option for "green" hydrogen production [5]. However, water electrolysis requires catalysts based partly on scarce or high-demand materials [6], and large quantities of (relatively expensive) electricity, motivating diversification of "green" hydrogen production technologies.

Photocatalysis offers a potentially cost-effective alternative by splitting water directly into hydrogen and oxygen based on (sun)light [7, 8], though it currently remains at lab scale [5]. Nevertheless, research suggests that photocatalysis can enable low levelized costs of hydrogen ($LCOH$) of around $1\,\$\,\mathrm{kg}_{H_2}^{-1}$ [9]. Studies [7, 8, 10–13] show that photocatalysis might compete with PV-powered electrolysis. These techno-economic analyses, however, overlook broader energy system effects such as sector interactions and flexibility.

This paper aims to bridge this gap by comparing photocatalytic hydrogen production with conventional electrolysis-based hydrogen production (mostly PV/wind-driven) on a system level. For this purpose, an energy system model is used to assess the impacts of integrating hydrogen from photocatalysis within the European energy system, enabling an analysis of the cost-optimal spatial system configuration and system operation. By varying the assumed cost of photocatalysis, different market shares of this technology are analyzed.

Within this context, the following key questions are assessed:

- How does the integration of hydrogen produced via photocatalysis influence the configuration of the European energy system compared to systems based on water electrolysis?
- In which geographical regions and under which conditions is photocatalysis most promising compared to water electrolysis?
- How might a shift in hydrogen production based on photocatalysis change the requirements for the energy-system related infrastructural needs?

## Photocatalysis

In photocatalytic systems, hydrogen and oxygen are produced directly by splitting water with sunlight via a photocatalyst (Fig. 1). This study assumes a particulate photocatalyst (particle size from nm to μm) suspended in water within simple large-scale flat baggie reactors (e.g., 300 × 12 m [7], 5 to 10 cm water level [8]). They are equipped with liquid inlets/outlets to supply the water/catalyst mixture and to remove the produced gas mixture. The hydrogen-oxygen mixture is produced by solar irradiation of the baggies, collected within them and is then transported to a central separation unit. Using suitable gas/gas separation methods such as pressure swing adsorption, pure hydrogen is obtained (Fig. 1). The separation energy demand is estimated to be ca. 10 % of the total energy produced (in the form of hydrogen) [7, 8]. Assuming that this energy demand is covered by self-produced hydrogen, raw hydrogen output must exceed plant design capacity by roughly 10 to 15 %. Finally, the remaining water/catalyst mixture is recycled (including possible treatment of the catalyst to restore activity) and cycled back into the baggie reactors (Fig. 1).

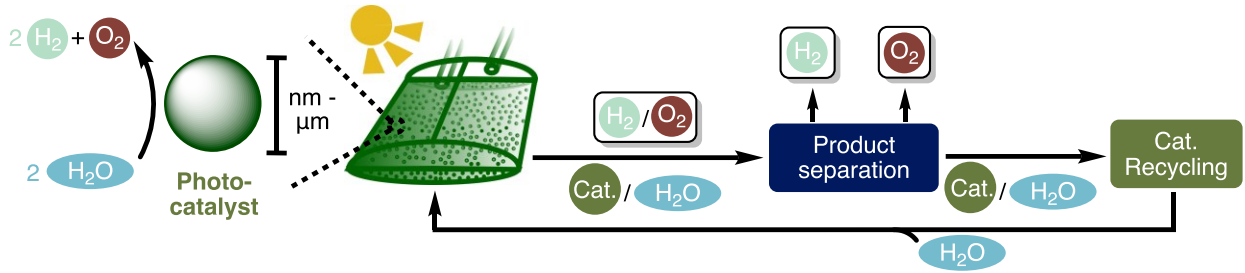

*Fig. 1: Photocatalysis process for the production of hydrogen, including photocatalyst particles splitting water into $H_2$ and $O_2$ inside the baggie reactor (left), separation of produced $H_2/O_2$ mixture (middle) and catalyst (Cat.) recycling (right).*

Such a photocatalytic hydrogen production, also referred to as type 1 photocatalysis system [7], has been found in techno-economic assessments to allow for the lowest-cost route to hydrogen production via solar water splitting (ca. 1.6 $ $kg_{H_2}^{-1}$, with a lower limit below 1 $ $kg_{H_2}^{-1}$ [7, 8]). Hydrogen production in baggie reactors has been demonstrated at lab-scale [14], and a related concept involving photocatalyst particles deposited on glass panels has been scaled up to 100 $m^2$, producing around 500 $L_{H_2}$ $d^{-1}$ [15].

Here, two metrics are particularly important to characterize this hydrogen production route: hydrogen production cost (varied to reach equal market share with electrolysis hydrogen) and solar-to-hydrogen ($STH$) efficiency. The latter quantifies the percentage of incident solar energy converted to hydrogen and determines required production area for a given amount of hydrogen. Experimentally, $STH$ efficiencies of up to 9 % have been demonstrated [16], although photocatalyst suitable for applications in baggie reactors are currently limited to around 2 % [17]. Theoretically, $STH$ efficiencies of 18 or 28 % (depending on the type of photocatalyst) are possible [18, 19], and research is ongoing [20].

Additionally, parameters, such as the required catalyst material, catalyst composition, recycling or precise energy demand and safety of hydrogen/oxygen separation are crucial for the deployment of photocatalysis systems [8]. However, they are not factored into the analysis presented here and are instead discussed qualitatively.

### Overview of system assumptions and case studies

For answering research our questions, physical, technical, economic, and political/legal assumptions must be made to define the system and constrain the decision space. General system assumptions used across all assessed cases are outlined below, followed by the parameter variations specific to each case.

The model PyPSA-Eur [21], for the optimization of capacity, operation and investment of various technologies (Section 3.1) is used. It optimizes the European energy system for a 96.8 % $CO_2$ emissions reduction by 2045 (baseline 1990), using 2040 cost projections to capture early investments essential for the desired system configuration. The system is clustered to 150 nodes and optimized using 3 h time steps based on 2013 weather data, balancing computational effort and memory requirements with resource availability [22, 23].

The model includes the heating, power, and industrial sectors, ensuring their respective and inelastic energy demands are met. Heat and electricity each account for roughly 40 % of the exogenously fixed energy demand and hydrogen/hydrogen-derivates for around 20 %. The transportation sector is excluded, as its energy demand may be met through energy imports, aligning with the EU's status as an energy importer. Imports or exports for hydrogen or electricity are not considered (fossil fuels utilization is limited by $CO_2$ emissions caps and limited carbon capture and storage). The hydrogen demand in the chosen sectors is heterogeneous, allowing for a detailed analysis of different hydrogen supply options.

Expandable generators for electricity and heat are solar thermal and photovoltaics (PV), onshore and offshore wind, open- and combined-cycle gas turbines (OCGT & CCGT), geothermal, biomass, and hydropower. Hydrogen is generated through electrolysis and photocatalysis. Existing oil, coal, and nuclear capacities may still be utilized. The model permits a 25 % expansion of the existing electricity transport grid, new hydrogen pipelines, and retrofitting existing gas pipelines for hydrogen transport. Storage options include batteries, short-term and long-term heat storage, and hydrogen storage in underground caverns (UGHS) or aboveground tanks. Methanol and oil demands are not spatially allocated, and biomass utilization is restricted to current levels. $CO_2$ sequestration is limited to $200 \, \text{Mt} \, \text{a}^{-1}$.

Photocatalysis reactors are installed with 0° slope and 180° azimuth. Maximum capacity per $\text{km}^2$ is based on typical photovoltaic values, adjusted to the efficiency of photocatalysis. At 10 % $STH$, the maximum installable capacity is $2.5 \, \text{MW}_{H_2} \, \text{km}^{-2}$; scaling with efficiency changes. Photocatalysis costs are expressed as annualized costs to flexibly represent any combination of fixed and operational expenditures, enabling adaptation of the model to different photocatalysis systems by adjusting costs, $STH$, and orientation.

To analyze the effects of photocatalysis implementation on the European energy system, photocatalysis costs are progressively reduced until photocatalytic hydrogen production reaches electrolysis output, targeting a balanced 50 % market share (equation (2)). Two edge cases are defined for the analysis:

1. **Electrolysis-only (PC-0)**: All hydrogen is produced via electrolysis.
2. **Balanced mix (PC-50)**: Hydrogen production is about evenly split between photocatalysis and electrolysis.

This study assumes a photocatalysis solar-to-hydrogen efficiency ($STH$) of 10 % and access to low-cost underground hydrogen storage (UGHS) where geologically feasible [24]. Fig. 2 shows the analyzed market shares over the annualized photocatalysis costs. Parameter variations for $STH$ efficiencies of 6 and 3 %, as well as limited storage are provided in Supplementary Note 1 and the market share results are shown in Supplementary Figure 1. The results of the analysis are presented based on the market shares.

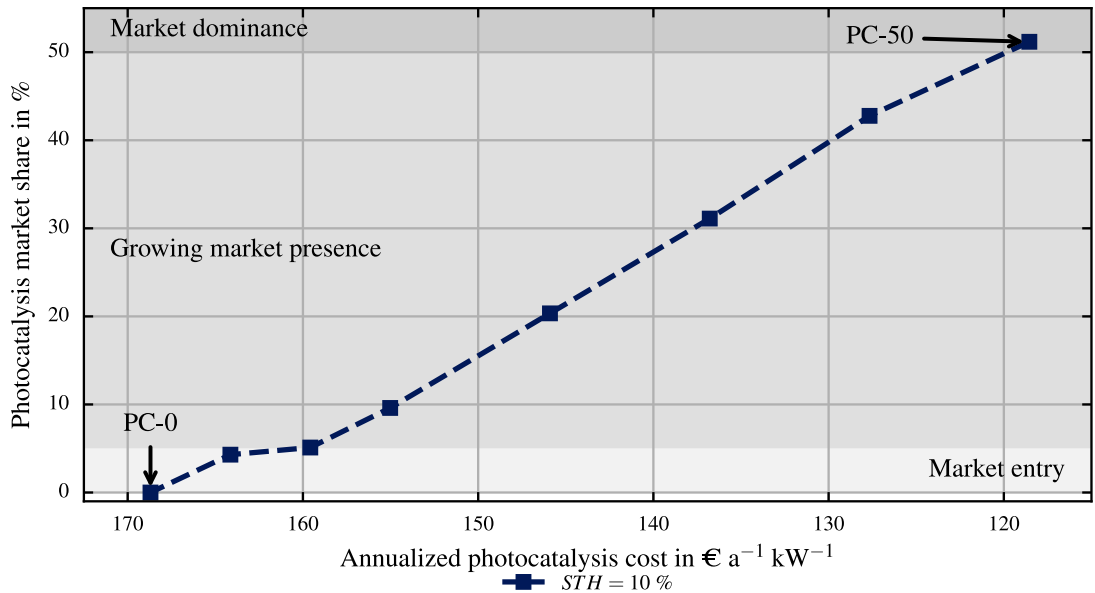

*Fig. 2: Market share analysis of photocatalysis (PC) relative to annualized photocatalysis costs, illustrating the linear trend with minor deviations for $10\,\%$ solar to hydrogen (STH) efficiency.*

### Hydrogen system development

The basis for the system analysis is the comparison of the case featuring hydrogen production solely by water electrolysis (PC-0 case) and a case combining hydrogen production by photocatalysis and electrolysis (PC-50 case).

Fig. 3 presents comparative maps of the spatial distribution and capacity deployment of hydrogen production and infrastructure in Europe. Important aspects of the PC-0 case are electrolysis capacity, hydrogen pipelines, and storage. Electrolysis is widely distributed, with a clear predominance in Western Europe, notably in the United Kingdom, Netherlands, Germany, and Denmark – nations bordering the North Sea. Hydrogen storage is located near electrolysis sites. Moderate electrolysis capacity is installed throughout France and Spain, while Eastern European countries show minimal installations (due to limited favorable renewable energy supply). A dense hydrogen pipeline network links the main electrolysis sites with Southern Europe. This layout aligns with [25].

In the PC-50 case, electrolysis and photocatalysis achieve a near-equal output of hydrogen. Here, electrolysis installations are reduced in northwestern Europe, while large photocatalysis installations have emerged in southern Europe, especially in Spain and Greece, with a smaller presence in Italy. Electrolysis is largely abandoned at photocatalysis sites. Large-scale storage capacities are installed near or at the photocatalysis sites. With a decreasing share of electrolysis, wind turbine and PV capacities decrease as well (Supplementary Note 2 and Supplementary Figure 2). While wind installations decrease at a similar rate to electrolysis, solar installations decrease at a higher rate; consequently, the total amount of generated electricity declines (reduced electrification) and hydrogen generation increases. Photocatalysis effectively substitutes a combination of PV, electrolysis, and wind installations.

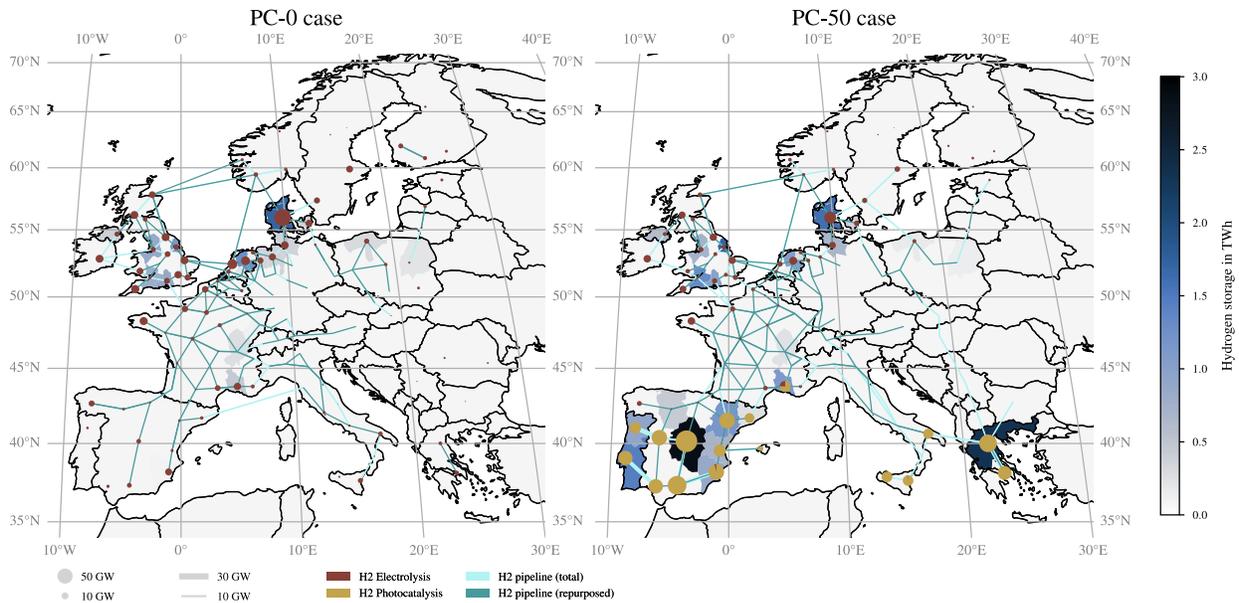

*Fig. 3: Comparison of the European hydrogen production and infrastructure in the electrolysis-only (PC-0) case and the balanced mix (PC-50) case with equal hydrogen production through electrolysis and photocatalysis.*

Furthermore, southern Europe's hydrogen pipeline network is significantly expanded, while Northern Europe's is slightly reduced (e.g., Scotland-Norway from two connections to one). New pipelines link photocatalysis sites without UGHS to nearby UGHS locations (e.g., southern Spain connected to central Spain). Overall, hydrogen production shifts from northern to southern Europe (similar to [26]).

The change in system design from the PC-0 to the PC-50 case can be attributed to the underlying renewable energy sources available for electrolysis and photocatalysis. Photocatalysis relies directly on solar irradiation, which is highly dependent on latitude. This explains the high number of South European installations. Additionally the strategic proximity of UGHS is crucial for optimizing energy storage and distribution (see Section 1.4) to tackle seasonal and inter-day fluctuations of solar irradiation. The storage enables loads to be temporally decoupled from the photocatalysis's day/night cycle.

In contrast, electrolysis plants can utilize multiple power sources and are concentrated near the North Sea, offering favorable wind conditions and diverse energy resources, making it one of Europe's optimal regions for electrolysis.

## Geographic shifts in $H_2$ supply and demand

Energy systems are complex and composed of interconnected generation, transport, storage, and demand technologies, each subjected to geographical, systemic, technical, economic, and further constraints. Energy generation from renewable sources is geographically constrained by weather patterns, whereas industrial activities and population density influence demand.

The installed capacities of electrolysis and photocatalysis at selected nodes are shown in Fig. 4. As photocatalysis gains market shares, hydrogen production shifts toward photocatalysis, particularly at southern nodes where a substantial decline in installed electrolysis is observed. The magnitude of capacity substitution varies regionally.

Many southern nodes quickly reach their maximum photocatalysis capacity and nearly all do so at the highest market share. Installations predominantly migrate from south to north with increasing market share, yet this trend is not solely latitude-dependent – some southern locations are developed later due to the lack of nearby UGHS. Photocatalysis installations are highest at the southernmost nodes with UGHS, where maximum expansion is quickly reached, while northern locations and those without UGHS are typically developed later (higher market shares).

Complementing the southward shift in production from PC-0 to PC-50 (Fig. 3 and Fig. 4), Fig. 5 shows a corresponding shift in hydrogen demand. As photocatalysis costs decrease and the respective market shares increase, hydrogen generation shifts southwards, with reduced hydrogen production by electrolysis and increased output by photocatalysis. Despite these changes, quartile positions shift only slightly, indicating stable technology-based production centers.

Hydrogen demand adjusts with the observed shift in production location: the median latitude of the demand shifts from ca. 52.5°N to ca. 42.5°N. Notable demand centers maintain their location, with northern areas between 50°N and 57.5°N remaining similar in shape across the analyzed market shares (although shrinking). The southern demand center expands northwards from around 38°N to 43°N. Conversely, large parts of Europe, like the region between 42.5°N and 50°N (Southern France to the Middle of Germany) show minimal hydrogen activity. High land area density (Fig. 5, right) doesn't necessarily equate to increased hydrogen usage, underscoring hydrogen demand alignment with generation locations rather than land density. In summary, the energy system's layout significantly shifts as photocatalysis gains prominence.

The flexibility in the hydrogen demand landscape is crucial for geographic supply shifts. While industrial hydrogen demand in this study, such as that for steel production (e.g., in the German Rhine-Ruhr area), is geographically fixed, the demand for hydrogen in methanol synthesis and Fischer-Tropsch processes used for producing chemical products and synthetic fuels is designed to be geographically independent. This allows for adaptation to generation patterns, enabling a southward shift in hydrogen demand (compare [26]) as market shares increase. This assumption is justified because the transportation costs for these fuels are relatively low, considering the current scale of energy carrier transport. Overall, the prominence of the North Sea area as a hydrogen production hotspot diminishes while new opportunities emerge in Southern Europe, notably in Spain, Italy, and Greece.

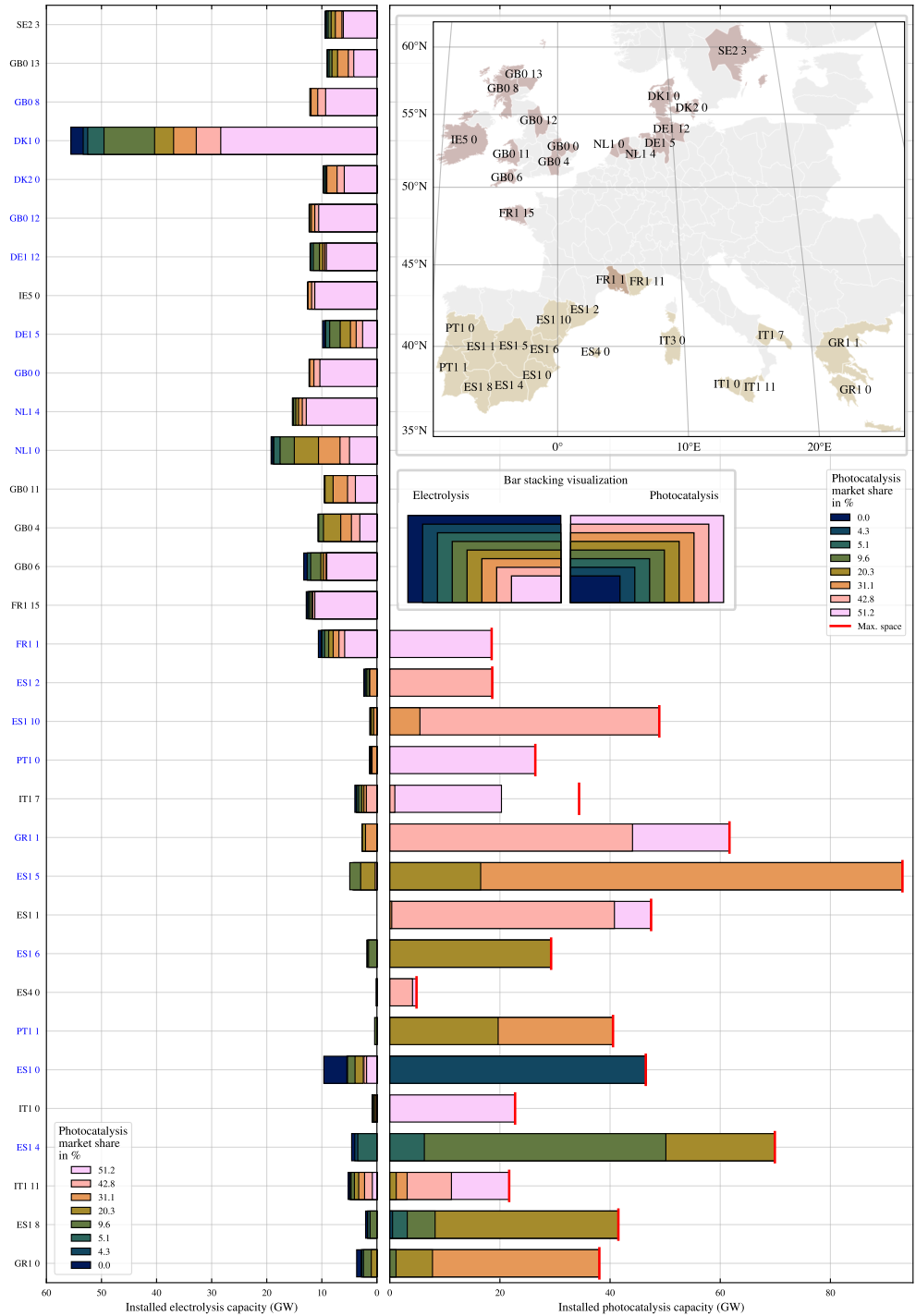

Fig. 4: Installed capacities of electrolysis and photocatalysis at selected nodes as overlaid bars: showing the top 18 nodes with significant photocatalysis and electrolysis installations (≥ 1 MW in the PC-50 case). Nodes are ordered from North to South (blue text is used for nodes where underground hydrogen storage is available). Electrolysis capacities are listed on the left, while photocatalysis is listed on the right. A map highlights nodes by predominant technology (dark red for electrolysis, ochre for photocatalysis). The red bar indicates the maximum space availability per node for photocatalysis installation.

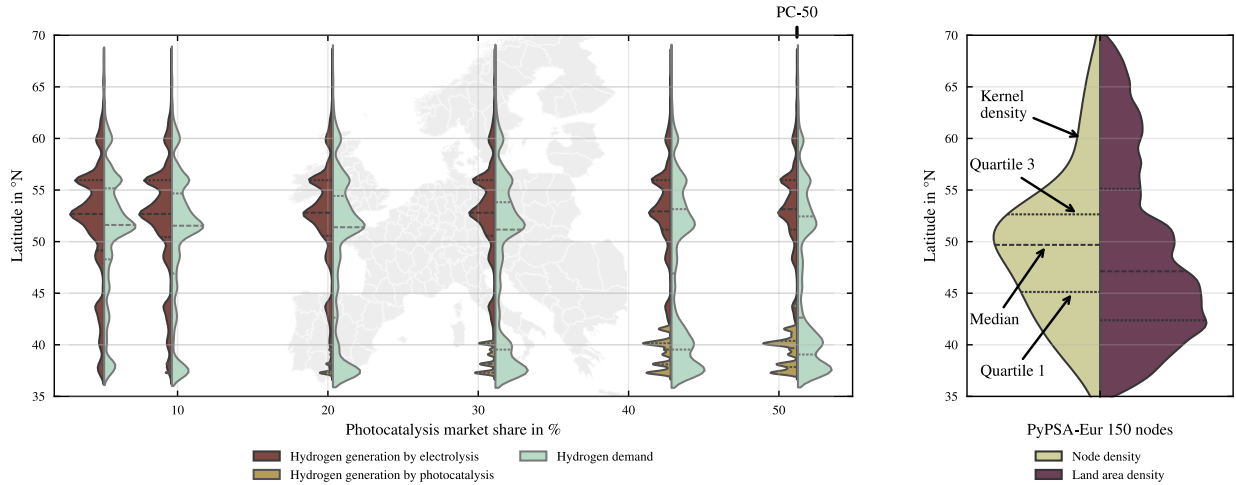

*Fig. 5: Violin plots of hydrogen generation and demand distribution over latitude for different photocatalysis market shares (left), node density, and land area density over latitude (right). The left plot shows generation by photocatalysis (ochre) and electrolysis (dark red) opposite to the demand (green), centered on key generation nodes. A map of Europe is added in the background for geographic reference regarding the latitude. (n = 150, weighted, bandwidth: Silverman, density norm: count)*

### Photocatalysis faces higher cost pressure than electrolysis

Fig. 6 compares the levelized cost of hydrogen (*LCOH*) for electrolysis and photocatalysis at varying photocatalysis market shares. Calculated nodally (equation (3) and equation (4)), the *LCOH* is displayed as a distribution. The leftmost violin represents the PC-0 case without photocatalysis, while the rightmost violin represents the PC-50 case. As photocatalysis installations increase and electrolysis installations decrease the *LCOH* distributions tend downwards, indicated by the median shift. For electrolysis, this trend can be attributed to a withdrawal from less favorable locations, enabling focus on optimal sites. For photocatalysis, the rise in market share (and exogenously lowered cost) is coupled with a requirement to use progressively less favorable sites due to space limitations (Section 1.4).

The median difference between both technologies ranges from 0.26 $€\,\text{kg}_{H2}^{-1}$ to 0.67 $€\,\text{kg}_{H2}^{-1}$. While the observable reduction in *LCOH* for both technologies can be attributed to the factors discussed, they do not fully account for the disparity in *LCOH* levels between electrolysis and photocatalysis. The persistent cost gap between the medians of photocatalysis and electrolysis shows that the systematic integration costs associated with photocatalysis exceed those of electrolysis.

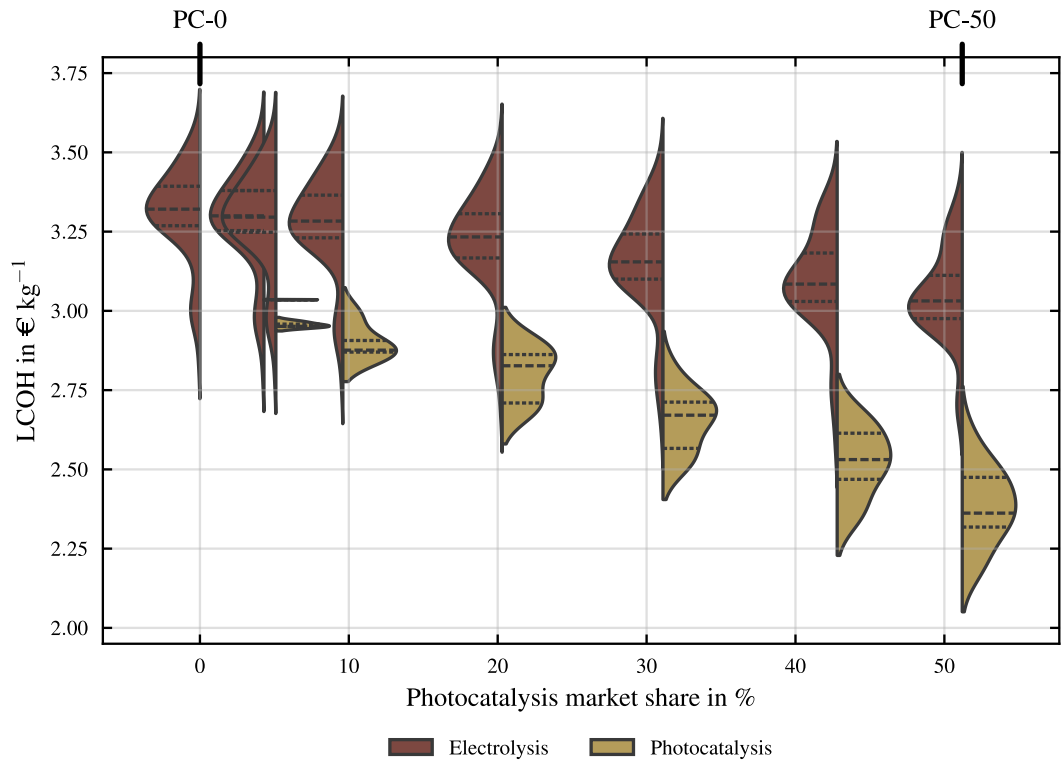

*Fig. 6: Comparison of levelized cost of hydrogen (LCOH) for electrolysis (dark red) and photocatalysis (ochre) across the energy system at various photocatalysis market shares. Refer to Supplementary Note 3 for an explanation why violins 2 & 3 are so close together. (n = 150, bandwidth: Silverman, density norm: width)*

## Photocatalysis integration alters energy system design

The configurations of the PC-0 and PC-50 cases are compared regarding several key system properties, as depicted in Fig. 7. The figure represents these differences in four thematic areas:

a) Electricity and hydrogen production
b) Hydrogen system
c) Heating and power system
d) Synthetic fuels

The bars in each block indicate differences in the yearly generation or total installed capacity on the left ordinate and in annualized costs in billion € a$^{-1}$ on the right ordinate. The PC-50 case shows a slight increase in hydrogen production and a corresponding reduction in electricity production. This trend is reflected in the associated costs. Photocatalysis substitutes some electrolysis and renewable energies such as solar and wind power. Additionally, there is an increase in natural gas utilization and a decrease in oil utilization (at the same $CO_2$ emission limit). These changes comprise the most significant portion of the cost changes.

The shift towards a greater hydrogen focus is underscored by the increased installation of hydrogen storage capacities and enhanced hydrogen transport capacity. However, these costs account for only about 0.10 € kg$_{H2}^{-1}$ of the *LCOH* difference (Fig. 6).

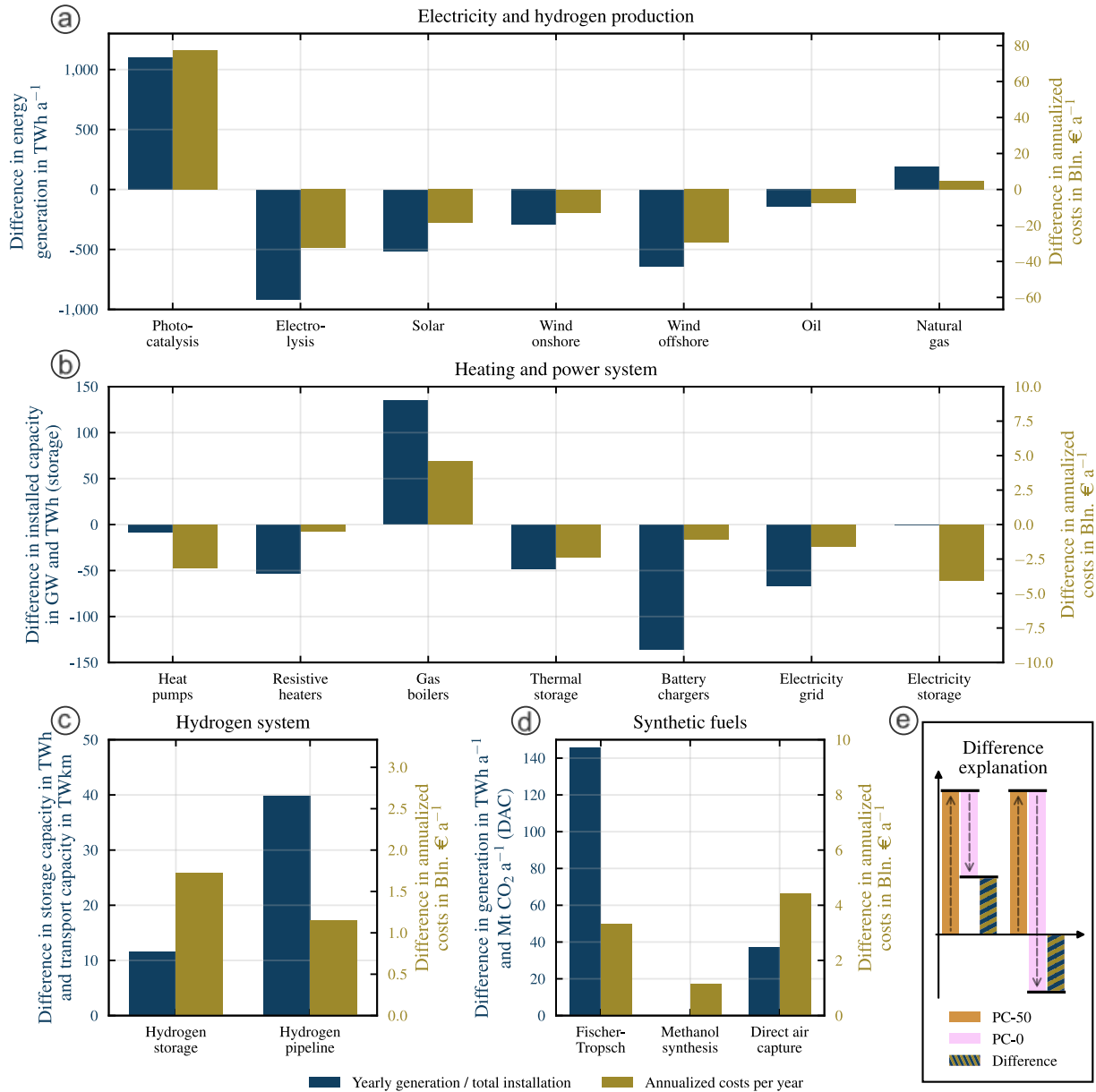

*Fig. 7: Capacity, generation, and cost differences between the PC-50 and PC-0 cases in key areas. Positive values represent increased installations and costs of the technology in the PC-50 case, characterized by significant photocatalysis utilization, whereas negative bars indicate reduced utilization compared to the PC-0 case (with electrolysis only).*

Another main change in the energy system configuration is the shift from high electrification towards gas-based technologies. This shift includes reductions in the installed heat pumps and resistive heaters, alongside decreased electrical and thermal storage capacities and a reduced electricity distribution grid. At the same time, the case shows higher gas boiler capacities. Additional hydrogen storage and pipeline infrastructure partially offset the reduced flexibility.

The shift towards less electricity utilization is further accompanied by the increase in the generation of synthetic fuels. Fischer-Tropsch fuel production is about 145 TWh a$^{-1}$ higher in the PC-50 case, necessitating an increase in direct air capture. Additional methanol synthesis capacities are added to provide flexibility for photocatalysis while the total amount of produced methanol remains the same.

In summary, an optimal European energy system with high penetration of photocatalysis exhibits less electrification but greater utilization of gas-based energy carriers compared to a system solely reliant on electrolysis for hydrogen production. Lower installations of electrolysis lead to a reduction in renewable electricity generation, since electrolysis provides a controllable load to synergize with the volatile electricity output from PV and wind. Synthetic fuels are utilized to compensate for the lower electrification. This increase in synthetic fuel generation reduces the need for fossil oil consumption compared to the PC-0 case. Consequently, the permissible net $CO_2$-emissions allocated to oil consumption in the PC-0 case can be used for natural gas consumption in the PC-50 case, providing flexibility in the power system.

## Conclusions

This study investigates the integration of photocatalytic hydrogen production into the European energy system. It examines how varying photocatalysis market shares affect energy system configuration across electricity, industry, and heating sectors in 150 regions with 3 h resolution for 2045, targeting low residual $CO_2$ emissions.

The analysis shows that photocatalysis cannot simply substitute electrolysis without significant system changes, as both technologies provide distinct functionality and flexibility. Electrolysis enables renewable electricity generation (wind, PV) and synergizes with electrification technologies like heat pumps and batteries, enhancing system resilience and flexibility. Photocatalysis must be around 20 % cheaper compared to electrolysis (in *LCOH*, depending on market share) to offset this flexibility advantage. Using photocatalysis results in a geographical southward shift of hydrogen production (toward higher solar irradiation), away from the North Sea area, an electrolysis hub due to its strong wind potential. This shift might lead to increased installation of hydrogen downstream industries (e.g., Fischer-Tropsch and methanol synthesis) in the southern Europe. High efficiency for photocatalysis is required to alleviate pressure on space demand, and inexpensive hydrogen storage possibilities are mandatory.

An optimal European energy system with a substantial amount of photocatalysis shows a different technology composition compared to an energy system without it, highlighting the need to consider photocatalysis early in infrastructure and demand planning to enable its potential and avoid lock-in effects. Although photocatalysis may reach significant deployment at future costs, it cannot fully replace electrolysis without altering system dynamics. However, its deployment diversifies "green" hydrogen production and can alleviate pressure on critical raw materials used, e.g., for PV [27] and electrolysis [6].

In conclusion, the potential of photocatalysis for diversifying large-scale hydrogen production depends on cost and efficiency improvements. Optimal deployment regions are in southern Europe with above-average solar irradiation. Successful integration of photocatalysis into energy systems requires strategic adjustments in system design, including infrastructure upgrades and demand planning.

## Methods

Below, energy system modeling using the PyPSA-Eur model is introduced, followed by a discussion of the implementation of photocatalysis in the energy system model. Finally, the assessment criteria for the influence of photocatalysis adoption on the European energy system are presented.

### Energy system model

The open-source model and dataset of the European energy system optimization model PyPSA-Eur [21] is used for capacity and deployment planning to meet demands while minimizing costs (i.e., the sum of annualized investments and operational costs) of the sector-coupled European energy system. PyPSA-Eur uses different openly available data sources to model the European energy infrastructure (e.g., electricity grid, gas grid), energy and energy carrier demands (e.g., electricity, heating), and generation technologies (i.e., power plants) and potentials (e.g., wind availability, solar availability) (Fig. 8). It includes the countries of the European Union (EU27), along with Norway, Switzerland, the United Kingdom, Bosnia and Herzegovina, Montenegro, Albania, North Macedonia, Serbia, and Kosovo, while Cyprus and Malta are excluded [21]. Regarding grid-based energy transport, this European region is considered isolated (i.e., no import or export is possible). PyPSA-Eur and the basic PyPSA model are validated e.g., in [21, 22, 28, 29]. The open-source approach allows a thorough analysis of the data and modeling foundation.

The energy system is represented as a graph with nodes and edges, where nodes correspond to regions with demand and supply, and edges represent the energy flow between the nodes. Inflexible demand curves for each sector are applied to the nodes using the spatial distribution of demand data and Voronoi tessellation. The model is designed for high spatial and temporal data resolution and constrained Linear Programming. Perfectly inelastic demands have to be met, and the model can determine the best-fitting energy system design and operation based on all available technologies to reach these demands. The most critical constraints are outlined below:

- Nodal energy balance, ensuring supply and demand match at each time step and node. Energy transmission to/from a node is supply, respectively, demand in that sense.
- $CO_2$ emission limits, restricting the utilization of energy from fossil resources.
- Weather-based availability for solar and wind resources.
- Spatial limitations on expanding renewable generators, based on protected sites and land classification.

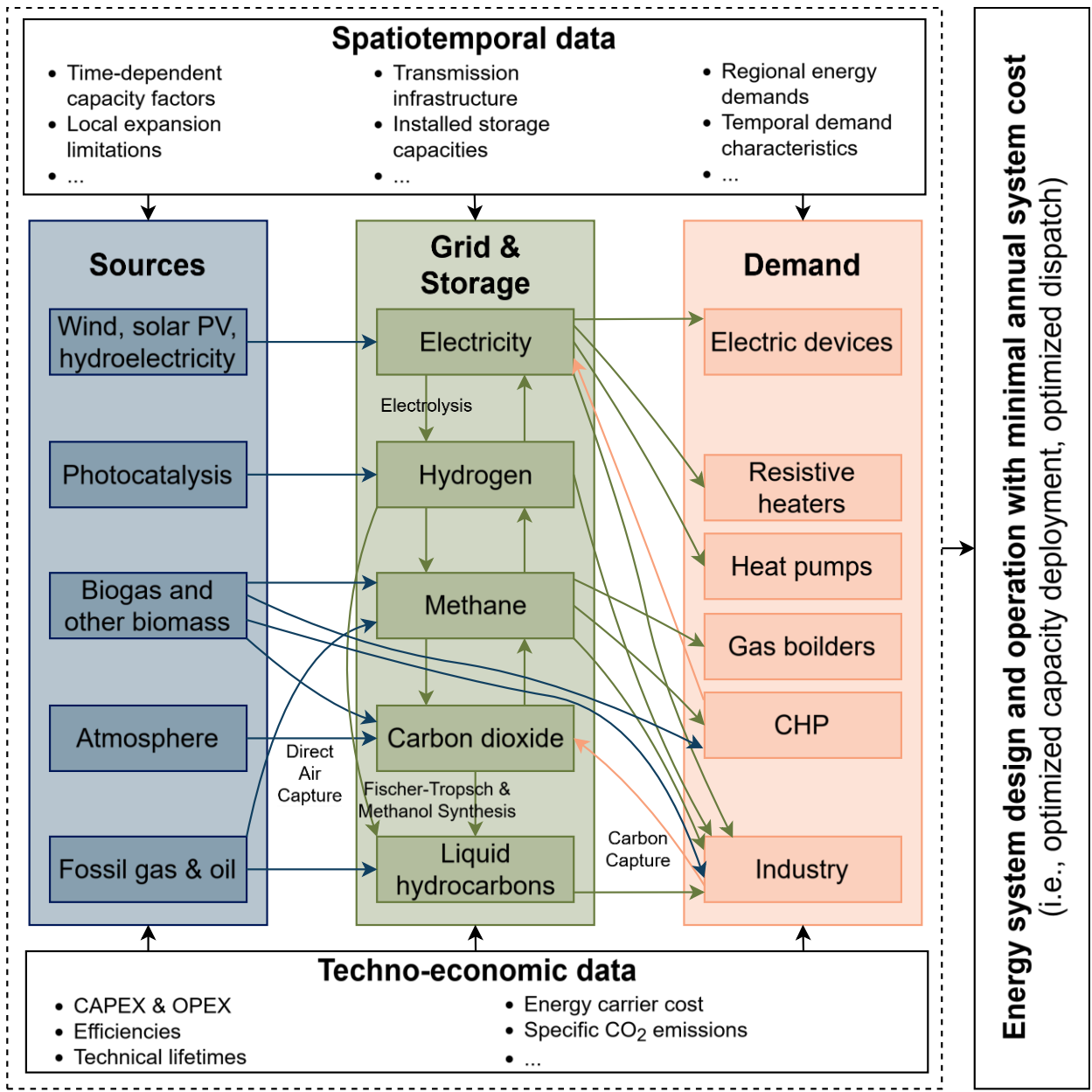

*Fig. 8: PyPSA-Eur model structure with added photocatalysis (adapted from [31]) (CHP: Combined heat and power).*

To yield meaningful results, considering the political, technological, and economic framework, large-scale energy system optimization models must be filled with large amounts of data for technologies, costs, and assumptions like future energy demands. Here, some fundamental to represent the European energy system is listed.

- Energy demand profiles are primarily from [32].
- Spatially and temporally resolved capacity factors for wind power, solar power (PV and thermal), run-off water for hydropower generation as well as heating and cooling demands are generated based on [33] using weather data from [34] and [35].

- Land-use restrictions are determined with land classifications based on [36] (excluding, e.g., infeasible areas like cities, bodies of water, natural reserves) and natural reserves data is extracted from [37].
- Grid infrastructure data stems from [38, 39].
- Technology data assumptions are taken from [40], with a large amount of data from [41].

## Model extensions and modifications

The current PyPSA-Eur model does not include photocatalysis. Thus this technology option is integrated into the model to investigate the systematic value and effects of integrating photocatalysis into the European energy system.

Photocatalysis is modelled as a device converting sunlight directly into hydrogen (water is considered to be always and everywhere available). The hydrogen output of the photocatalysis systems ($\dot{V}$) is calculated based on the total solar irradiation on a tilted plane ($I$), the solar-to-hydrogen efficiency ($STH$) of the respective device, and the properties of hydrogen and water (equation (1)). Specifically, the molar mass ($M_{H_2}$) and the lower heating value of hydrogen ($LHV_{H_2}$), as well as the Gibbs free energy change for water splitting at standard conditions ($\delta G_{r0}$) are used. Capacity factors are derived by dividing the output by a reference output at solar irradiation of 1000 W m$^{-2}$.

$$\dot{V}(STH, irradiance) = \frac{STH \cdot I}{\delta G_{r0}} \cdot M_{H_2} \cdot LHV_{H_2} \quad (1)$$

The following techno-economic data can be adjusted for the implemented photocatalysis model:

- solar to hydrogen efficiency ($STH$)
- investment costs (CAPEX)
- operational costs (OPEX)
- technical lifetime
- irradiation data
- system orientation (slope and azimuth)
- spatial expansion limit (in hydrogen production capacity per km²)
- land classification types (land types photocatalysis can be built on)

Photocatalysis systems are implemented as a novel hydrogen generator being connected to the hydrogen bus to directly produce hydrogen from weather-dependent solar energy. Such systems are subject to output restrictions to nominal power and weather-dependent capacity factors and generator expansion limits based on land classification, natural reserves and restrictions to the maximum installations per area. The addition of photocatalysis provides the energy system model PyPSA-Eur with an extra degree of freedom regarding generator choice to meet sectoral demands (for additional information see Supplementary Note 4).

## Assessment criteria

Qualitative and quantitative comparisons are conducted to evaluate the optimization results. The electrolysis-only system (PC-0) and the balanced mix system (PC-50) are discussed qualitatively by examining the spatial distribution and capacity deployment of hydrogen production technologies in Europe. The systematic impact of integrating photocatalysis systems within the European energy system is analyzed including the examination of the geographic distribution and shifts in hydrogen supply and demand based on photocatalysis-based hydrogen provision costs. The assessment is focused on aggregated geospatial data, incorporating kernel density estimations (i.e., determination of an unknown probability function), and evaluation of economic impacts through the $LCOH$ and comparing system cost differences among selected technologies.

### Market share

To quantitatively assess the hydrogen production contribution of photocatalysis relative to water electrolysis, the market share ($S_{PC,EL}$) is calculated. This metric is the ratio of hydrogen produced by photocatalysis systems to the overall hydrogen output at all nodes (equation (2)). $g_{n,PC,t}$ and $g_{n,EL,t}$ is the hourly hydrogen production from photocatalysis and electrolysis at node $n$.

$$S_{PC,EL} = \frac{\sum_{n=0}^{N} \sum_{t=0}^{T} g_{n,PC,t}}{\sum_{n=0}^{N} \left( \sum_{t=0}^{T} g_{n,EL,t} + \sum_{t=0}^{T} g_{n,PC,t} \right)} \quad (2)$$

### Levelized cost of hydrogen

The levelized costs are an essential metric for comparing the economic viability of production facilities. Here, the levelized costs of hydrogen ($LCOH$) are calculated for electrolysis and photocatalysis to compare the two technologies.

The $LCOH$ for electrolysis at node $n$ ($LCOH_{EL,n}$) is determined by dividing the total costs of hydrogen production by the aggregated hydrogen production (equation (3)). To calculate the cost of hydrogen production for each time step $t$, the nodal electricity price $\lambda_{n,t}$ is multiplied by nodal electricity demand for electrolysis $D_{n,EL,t}$, weighted per time step $w_t$. Then, the fixed annualized cost per electrolysis capacity ($c_{n,EL}$) multiplied by the installed capacity at node $n$ ($G_{n,EL}$) are added. The resulting value is then divided by the sum of the hourly hydrogen production from electrolysis at node $n$ $g_{n,EL,t}$, weighted per time step $w_t$.

$$LCOH_{EL,n} = \frac{\sum_{t=0}^{T} (\lambda_{n,t} \, D_{n,EL} \, w_t) + c_{n,EL} \, G_{n,EL}}{\sum_{t=0}^{T} g_{n,EL,t} \, w_t} \quad (3)$$

For hydrogen production via photocatalysis, which does not rely on an external power supply, the $LCOH$ calculation simplifies to the annualized costs of photocatalysis ($c_{n,PC}$), the installed capacity at the node $n$ ($G_{n,PC}$) and the nodal hydrogen generation from photocatalysis ($g_{n,PC,t}$).

$$LCOH_{PC,n} = \frac{c_{n,PC} \, G_{n,PC}}{\sum_{t=0}^{T} g_{n,PC,t} \, w_t} \quad (4)$$

## Acknowledgements

W.T gratefully acknowledges financial support by the DAAD with funds from the Federal Foreign Office (AA) through the project "Renewable Energies with 100 % Guaranteed Power Supply" as part of the Ta'ziz program (project number 57682003).

J.S. gratefully acknowledges financial support by Fonds der Chemischen Industrie (Liebig scholarship for J.S.), Federal Ministry of Research, Technology and Space (BMFTR independent research group "SINATRA: SolSTEP", grant number 03SF0729) and DFG via the CRC TRR 234 CATALIGHT (project no. 36454990, project A7).

## Data availability

The code for this study and all associated data is available on GitHub and and zenodo. Assessment and plotting: https://github.com/w-tusche/photocatalysis-europe; changes to PyPSA-Eur: https://github.com/w-tusche/pypsa-eur-pc; changes to atlite: https://github.com/w-tusche/atlite-pc; raw data of PC-0 and PC-50 cases https://doi.org/10.5281/zenodo.16360844.

## Supplementary Material

### Supplementary Note 1: Market shares & Parameter variation

To analyze the effects of photocatalysis implementation in the European energy system, the threshold cost for photocatalysis deployment is determined by exogenous changes. The threshold cost is where photocatalysis is not economically viable, but a slight cost reduction leads to first installations. It is determined iteratively since the share of photocatalysis in the energy system is determined endogenously. From the threshold value on, the cost of photocatalysis (capex in the costs_2040.csv) is reduced exogenously for each optimization in discrete steps until the share of hydrogen from photocatalysis is higher than that of electrolysis (balanced mix with about 50 % $H_2$ from photocatalysis). The iterative nature and the discrete change are the reason why the PC-50 case is does not represent a perfect 50:50 split.

The analysis in the main text of the study focussed on results with $STH = 10\%$. The same tendencies and mechanisms can be observed for all other analyzed parameter combinations, however with an offset in the total installed photocatalysis (compare Supplementary Figure 1). Additionally to the cases shown in the figure, the two boundary cases (PC-0 and PC-50) were analyzed with a net zero $CO_2$ emissions constraint with similar system configurations.

The shift from a system with electrolysis only to increased photocatalysis installation was exogenously influenced by adjusting the annualized photocatalysis cost to analyze the thresholds at which photocatalysis becomes competitive and to identify optimal installation locations.

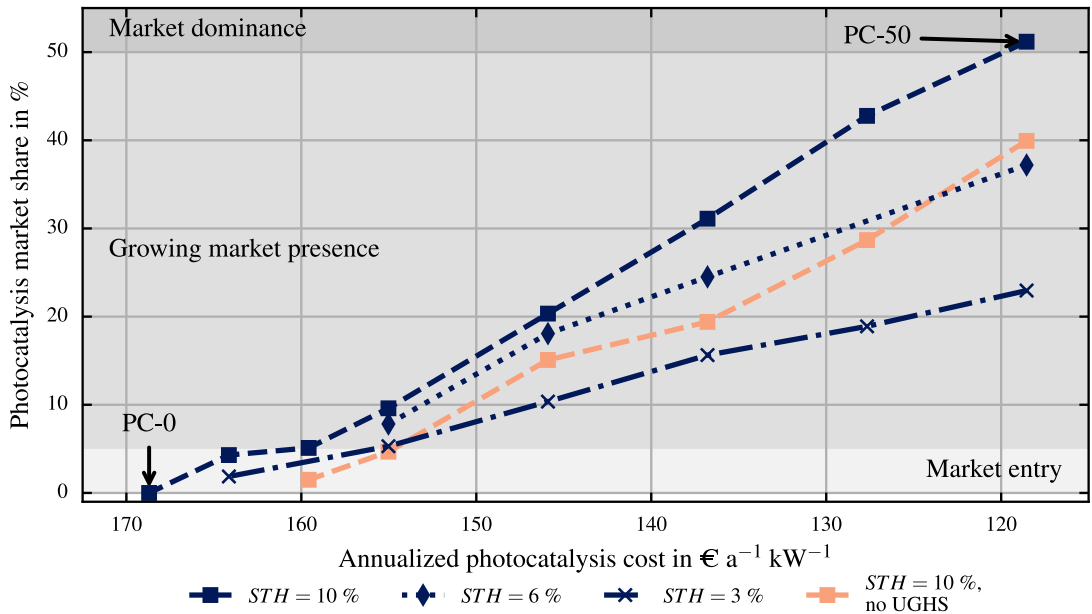

*Supplementary Figure 1: Market share analysis of photocatalysis relative to annualized photocatalysis costs, illustrating the linear trend with minor deviations across various solar to hydrogen (STH) efficiencies (3, 6, and 10 %) and a case with 10 % adn no low-cost underground hydrogen storage (UGHS).*

The market shares for photocatalysis as a function of annualized cost have a nearly linear trend with minor deviations. The trend is consistent across all tested efficiencies (3, 6, and 10 %) and in the case with 10 % efficiency and no available underground hydrogen storage (UGHS) as a low-cost hydrogen storage option.

The availability of low-cost cavern storage for hydrogen shows a similar effect on the market share of photocatalysis within the energy system as reducing the photocatalysis efficiency from 10 % to 6 %, outlining the significance of low-cost hydrogen storage. All curves begin at a similar point in the lower left corner of the figure, but as annualized photocatalysis costs decrease, the curves for different efficiencies diverge. There is an approximate 25 % reduced market share for the 6 % efficiency case and a 50 % reduction for the 3 % efficiency case compared to the balanced mix case with a 10 % efficiency.

Overall, changes in efficiency, at a constant annualized photocatalysis cost, have a more pronounced impact as installations increase. This effect can be attributed to spatial limitations in regions with high solar irradiation. Lower efficiency photocatalysis necessitates more space for the same hydrogen output (lower power per area ($kW\,m^{-2}$)), yet installation is constrained at the nodes with the best photocatalysis potential, forcing the system to install photocatalysis in less favorable locations.

General observations form the parameter variations are:

- Low-cost hydrogen storage plays a major role to enable large-scale photocatalysis deployment.

- Photocatalysis is installed from South to North until space per node is no more available. Lower STH values allow for less installed power per node, forcing photocatalysis installations northwards and making installation of photocatalysis unfavorable.

## Supplementary Note 2: Energy system development

A detailed picture of the changes in the deployment of technologies is depicted in Supplementary Figure 2, which shows the change in installed capacity and energy generation for selected components as annualized photocatalysis costs alternate.

In the electrolysis-only system (PC-0), photocatalysis is economically unviable, with costs of ca. 169 € $a^{-1}$ $kW^{-1}$; photocatalysis systems are not installed by the model under these conditions. Conversely, photocatalysis has reached substantial deployment to produce about half of all hydrogen in the PC-50 case at ca. 118 € $a^{-1}$ $kW^{-1}$. The data points between these extremes show the transition path. Additionally, two more axes are provided to compare against other potentially relevant cost relations, namely the ratios between photocatalysis and photovoltaics and electrolysis costs. A change of any technology cost inside the model can change the system layout. If, e.g., electrolysis costs or photovoltaic costs are reduced, photocatalysis installation becomes less favorable, and lower costs of photocatalysis are required for the same number of installations. Generally, the technology balances are of course cost-dependent.

Supplementary Figure 2 shows the installed generation capacity for photocatalysis, electrolysis, and renewable energies solar and on-/offshore wind. Furthermore, it shows the annual energy generation for hydrogen and electricity. For electricity, Supplementary Figure 2 only considers the production from primary energy sources, thereby excluding secondary electricity generation from stored energy to provide a picture of the initial energy input.

As the system changes from PC-0 to PC-50, the installed generation capacity for electrolysis decreases while photocatalysis installations increase. In parallel with the decrease in electrolysis the amount of installed wind turbine and PV capacity decreases. While wind installations decrease at a similar rate as electrolysis, solar installations decrease at a higher rate. Consequently, the total amount of electricity generated also declines. On the contrary, with increasing photocatalysis installations, the overall generation of hydrogen increases in the energy system.

In summary, the change from the PC-0 to the PC-50 system progresses nearly linearly concerning the declining photocatalysis costs across all selected technologies, although at varying growth rates. As the integration of photocatalysis increases, the proportion of energy derived from hydrogen rises, and the total electricity generation diminishes significantly. This trend suggests a change towards reduced electrification. Photocatalysis effectively substitutes a combination of solar and electrolysis installations and wind to a lesser extent.

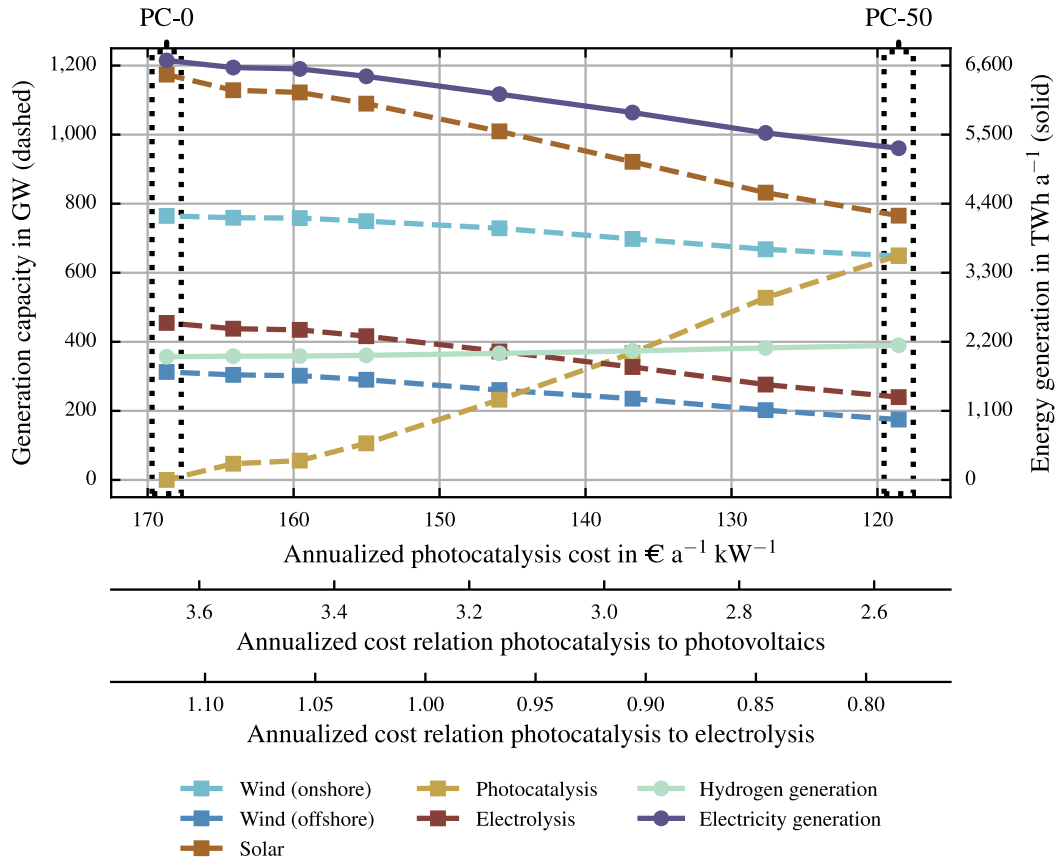

*Supplementary Figure 2: Change in overall system development as a function of annualized photocatalysis costs. The figure demonstrates the variation in generation capacity (GW) for photocatalysis, electrolysis, solar, and wind energy (onshore and offshore), as well as the energy generation for hydrogen and electricity. PC-0 represents the electrolysis only system without any photocatalysis, PC-50 a system where half of the hydrogen is produced via photocatalysis.*

## Supplementary Note 3: Non linearity in the market share annualized photocatalysis costs curve

The market share increase per decrease in annualized photocatalysis cost has for the most part the same (or similar) rate of change, however in Fig. 2 between 164.1 € $a^{-1}$ kW$^{-1}$ and 159.6 € $a^{-1}$ kW$^{-1}$ is reduced. The reason for this is that one location (node) in the South of Spain has conditions for photocatalysis that are a lot better than all other locations. Thus, at node ES1 0 photocatalysis is installed at comparatively high photocatalysis costs, while for all other nodes the costs have to be reduced further. The jump results from ES1 0 being fully developed right away at 164.1 € $a^{-1}$ kW$^{-1}$, while installations at all other nodes only start at a small rate at 159.6 € $a^{-1}$ kW$^{-1}$. This can also be observed well in Fig. 4 and Fig. 6 where the high ochre bar in the second violin from the left represents the installations at node ES1 0 and only a narrow photocatalysis violin in the third violin. Due to the comparatively large jump in annualized photocatalysis cost required and the study design by changing annualized photocatalysis costs (exogenous value) rather than market shares (endogenous value, which would require an iterative process) the two violins in Fig. 6 are that close.

## Supplementary Note 4: PyPSA and atlite model extensions

As a basis for the inclusion of photocatalysis into the the optimization model PyPSA 0.11.0 and atlite v0.2.12 are used and adjusted accordingly.

To include photocatalysis inside PyPSA-Eur it is added to atlite and inside PyPSA as a generator (inheriting all generator constraints). Inside PyPSA-Eur the customized atlite submodule is used to calculate capacity factors. The implementation in atlite is oriented on the implementation of photovoltaic. In PyPSA-Eur the maximum expansion limits are loosely based on the value of solar expansion limits: 5.1 MW at about 20 % efficiency for solar and 2.5 MW at about 10 % STH of photocatalysis. This is done since area of installation is assumed for the two technologies. The maximum expansion limit is based on many factors and can change based on the social acceptance and political frame work conditions. Through the change of efficiency a change of expansion limit was indirectly considered in the study. It is shown, that the expansion limit greatly influences the implementation of photocatalysis.

## Literature

1. IPCC (2023) Climate Change 2023: Synthesis Report - A Report of the Intergovernmental Panel on Climate Change. Contribution of Working Groups I, II and III to the Sixth Assessment Report of the Intergovernmental Panel on Climate Change [Core Writing Team, H. Lee and J. Romero (eds.)]. Intergovernmental Panel on Climate Change (IPCC), Geneva, Switzerland

2. Gielen D, Boshell F, Saygin D, et al. (2019) The role of renewable energy in the global energy transformation. Energy Strategy Reviews 24:38–50. doi: 10.1016/j.esr.2019.01.006

3. Ramsebner J, Haas R, Ajanovic A, Wietschel M (2021) The sector coupling concept: A critical review. WIREs Energy and Environment 10:e396. doi: 10.1002/wene.396

4. Muscat A, de Olde EM, de Boer IJM, Ripoll-Bosch R (2020) The battle for biomass: A systematic review of food-feed-fuel competition. Global Food Security 25:100330. doi: 10.1016/j.gfs.2019.100330

5. IEA (2024) Global Hydrogen Review 2024. International Energy Agency (IEA)

6. Eikeng E, Makhsoos A, Pollet BG (2024) Critical and strategic raw materials for electrolysers, fuel cells, metal hydrides and hydrogen separation technologies. International Journal of Hydrogen Energy 71:433–464. doi: 10.1016/j.ijhydene.2024.05.096

7. Pinaud BA, Benck JD, Seitz LC, et al. (2013) Technical and economic feasibility of centralized facilities for solar hydrogen production via photocatalysis and photoelectrochemistry. Energy & Environmental Science 6:1983. doi: 10.1039/c3ee40831k

8. Schneidewind J (2022) How Much Technological Progress is Needed to Make Solar Hydrogen Cost-Competitive? Advanced Energy Materials. doi: 10.1002/aenm.202200342

9. James BD, Baum GN, Perez J, Baum KN (2009) Technoeconomic Analysis of Photoelectrochemical (PEC) Hydrogen Production. doi: 10.2172/1218403

10. Shaner MR, Atwater HA, Lewis NS, McFarland EW (2016) A comparative technoeconomic analysis of renewable hydrogen production using solar energy. Energy & Environmental Science 9:2354–2371. doi: 10.1039/C5EE02573G

11. Grimm A, Jong WA, Kramer GJ (2020) Renewable hydrogen production: A techno-economic comparison of photoelectrochemical cells and photovoltaic-electrolysis. International Journal of Hydrogen Energy 45:22545–22555. doi: 10.1016/j.ijhydene.2020.06.092

12. Grube T, Reul J, Reuß M, et al. (2020) A techno-economic perspective on solar-to-hydrogen concepts through 2025. Sustainable Energy & Fuels 4:5818–5834. doi: 10.1039/D0SE00896F

13. Frowijn LSF, van Sark WGJHM (2021) Analysis of photon-driven solar-to-hydrogen production methods in the Netherlands. Sustainable Energy Technologies and Assessments 48:101631. doi: 10.1016/j.seta.2021.101631

14. Zhao Z, Goncalves RV, Barman SK, et al. (2019) Electronic structure basis for enhanced overall water splitting photocatalysis with aluminum doped $SrTiO_3$ in natural sunlight. Energy & Environmental Science 12:1385–1395. doi: 10.1039/C9EE00310J

15. Nishiyama H, Yamada T, Nakabayashi M, et al. (2021) Photocatalytic solar hydrogen production from water on a 100-M2 scale. Nature 598:304–307. doi: 10.1038/s41586-021-03907-3

16. Zhou P, Navid IA, Ma Y, et al. (2023) Solar-to-hydrogen efficiency of more than 9% in photocatalytic water splitting. Nature 613:66–70. doi: 10.1038/s41586-022-05399-1

17. Liu J, Liu Y, Liu N, et al. (2015) Metal-free efficient photocatalyst for stable visible water splitting via a two-electron pathway. Science 347:970–974. doi: 10.1126/science.aaa3145

18. Fountaine KT, Lewerenz HJ, Atwater HA (2016) Efficiency limits for photoelectrochemical water-splitting. Nature communications 7:13706. doi: 10.1038/ncomms13706

19. Schneidewind J, Argüello Cordero MA, Junge H, et al. (2021) Two-photon, visible light water splitting at a molecular ruthenium complex. Energy & Environmental Science 14:4427–4436. doi: 10.1039/D1EE01053K

20. Gunawan D, Zhang J, Li Q, et al. (2024) Materials Advances in Photocatalytic Solar Hydrogen Production: Integrating Systems and Economics for a Sustainable Future. Advanced Materials 2404618. doi: 10.1002/adma.202404618

21. Neumann F, Zeyen E, Victoria M, Brown T (2023) The potential role of a hydrogen network in Europe. Joule 7:1793–1817. doi: 10.1016/j.joule.2023.06.016

22. Frysztacki MM, Hörsch J, Hagenmeyer V, Brown T (2021) The strong effect of network resolution on electricity system models with high shares of wind and solar. Applied Energy 291:116726. doi: 10.1016/j.apenergy.2021.116726

23. Schlachtberger DP, Brown T, Schäfer M, et al. (2018) Cost optimal scenarios of a future highly renewable European electricity system: Exploring the influence of weather data, cost parameters and policy constraints. Energy 163:100–114. doi: 10.1016/j.energy.2018.08.070

24. Caglayan DG, Weber N, Heinrichs HU, et al. (2020) Technical potential of salt caverns for hydrogen storage in Europe. International Journal of Hydrogen Energy 45:6793–6805. doi: 10.1016/j.ijhydene.2019.12.161

25. Neumann F, Brown T (2021) The near-optimal feasible space of a renewable power system model. Electric Power Systems Research 190:106690. doi: 10.1016/j.epsr.2020.106690

26. Hofmann F, Tries C, Neumann F, et al. (2025) H2 and CO2 network strategies for the European energy system. Nature Energy 10:715–724. doi: 10.1038/s41560-025-01752-6

27. Zhang Y, Kim M, Wang L, et al. (2021) Design considerations for multi-terawatt scale manufacturing of existing and future photovoltaic technologies: Challenges and opportunities related to silver, indium and bismuth consumption. Energy & Environmental Science 14:5587–5610. doi: 10.1039/D1EE01814K

28. Brown T, Hörsch J, Schlachtberger D (2018) PyPSA: Python for Power System Analysis. Journal of Open Research Software 6:4. doi: 10.5334/jors.188

29. Hörsch J, Hofmann F, Schlachtberger D, Brown T (2018) PyPSA-Eur: An open optimisation model of the European transmission system. Energy Strategy Reviews 22:207–215. doi: 10.1016/j.esr.2018.08.012

30. Lange J, Schulthoff M, Puszkiel J, et al. (2024) Aboveground hydrogen storage – Assessment of the potential market relevance in a Carbon-Neutral European energy system. Energy Conversion and Management 306:118292. doi: 10.1016/j.enconman.2024.118292

31. Victoria M, Zeyen E, Brown T (2022) Speed of technological transformations required in Europe to achieve different climate goals. Joule 6:1066–1086. doi: 10.1016/j.joule.2022.04.016

32. Rozsai M, Jaxa-Rozen M, Salvucci R, et al. (2024) Integrated Database of the European Energy System (JRC-IDEES).

33. Hofmann F, Hampp J, Neumann F, et al. (2021) Atlite: A Lightweight Python Package for Calculating Renewable Power Potentials and Time Series. Journal of Open Source Software 6:3294. doi: 10.21105/joss.03294

34. Hersbach H, Bell B, Berrisford P, et al. (2023) ERA5 hourly data on single levels from 1940 to present. doi: 10.24381/CDS.ADBB2D47

35. Pfeifroth U, Kothe S, Müller R, et al. (2017) Surface Radiation Data Set - Heliosat (SARAH) - Edition 2. 7.1 TiB. doi: 10.5676/EUM_SAF_CM/SARAH/V002

36. European Environment Agency (2019) CORINE Land Cover 2018 (raster 100 m), Europe, 6-yearly - version 2020_20u1, May 2020. doi: 10.2909/960998C1-1870-4E82-8051-6485205EBBAC

37. European Environment Agency, European Commission (2024) Natura 2000 (vector) - version 2022. doi: 10.2909/95E717D4-81DC-415D-A8F0-FECDF7E686B0

38. ENTSO-E (2024) Grid Map. ENTSO-E Transmission System Map

39. Wiegmans B (2016) GridKit: GridKit 1.0 'for Scientists'. doi: 10.5281/ZENODO.47263

40. lisazeyen, euronion, Neumann F, et al. (2024) PyPSA/technology-data: V0.9.0. doi: 10.5281/ZENODO.11181492

41. DEA (2024) Analyses and statistics  Energistyrelsen. Danish Energy Agency